\documentstyle[prl,aps,epsfig,twocolumn]{revtex}
\begin{document} 
\draft \flushbottom \twocolumn[
\hsize\textwidth\columnwidth\hsize\csname @twocolumnfalse\endcsname

\title{Current instability and diamagnetism in small-diameter carbon
nanotubes}
\author{J. Gonz\'alez \\}
\address{
        Instituto de Estructura de la Materia.
        Consejo Superior de Investigaciones Cient{\'\i}ficas.
        Serrano 123, 28006 Madrid. Spain.}

\date{\today}
\maketitle
\begin{abstract}
\widetext
We investigate the electronic instabilities in carbon
nanotubes of short radius, looking for the breakdown of the
Luttinger liquid regime from the singular behavior of the
charge stiffnesses at low energies. We show that such a
breakdown is realized in the undoped (3,3) nanotubes through 
the onset of phase separation into regions with opposite 
electronic current. The phenomenology derived from this 
regime is consistent with the formation of a pseudogap in 
the single-particle spectrum as well as with a divergent
diamagnetic susceptibility, as observed in the experiments
carried out in carbon nanotubes of small diameter.

\end{abstract}

]

\narrowtext 
\tightenlines


Carbon nanotubes (CN) are among the best candidates for the
development of devices in molecular electronics. For this
reason, there has been great interest during the last years in
the investigation of their electronic properties. As the 
electrons are constrained to move in the reduced dimensions
of the tubule, CN constitute a paradigm of strongly correlated 
electron system\cite{bal,eg,kane,yo}. 
Thus, it has been possible to observe signatures of Coulomb 
blockade in short nanotube samples, as well as signatures of 
1D transport characteristic of Luttinger 
liquid behavior\cite{exp,yao}.

Yet there has been also clear evidence of superconducting (SC)
correlations in CN attached to suitable contacts\cite{kas,marc}. 
Supercurrents have been observed in the samples with SC 
electrodes reported in Ref. \onlinecite{kas}, providing evidence 
of the proximity effect in the CN\cite{th1}. 
Moreover, SC transitions have been measured in nanotube ropes 
attached to highly transparent contacts\cite{sup}. More
recently, measurements carried out in CN inserted
in a zeolite matrix have supported the conclusion that strong
SC correlations should exist in CN of very
short diameter, leading to a 
transition temperature $T_c \sim 15 \; {\rm K}$ \cite{chi}.

The experiment reported in Ref. \onlinecite{chi} 
has provided evidence of the diamagnetic behavior
of the CN at low temperatures. From inspection
of the $I$-$V$ characteristics, it has been also appreciated the
appearance of a gap in the single-particle spectrum, which has
been used to infer the value of $T_c$ \cite{chi}.
However, the lack of evidence for a true SC transition
stresses the fact that the mentioned observations
refer to properties of the individual CN, which cannot be
coupled by electron hopping as they only have a weak interaction
with the zeolite walls\cite{chi}. In general,
the tunneling amplitude between metallic 
nanotubes is what dictates the setting of 3D Cooper-pair 
coherence and the value of $T_c$ \cite{th2,th3}. 
The existence of a low-energy scale intrinsic to the 
individual CN demands therefore a reexamination of the 
role of the dominant correlations in nanotubes of short
radius.

The nanotube diameter $d = 4.2 \pm 0.2 {\rm \AA}$ reported in
Ref. \onlinecite{chi} is closer to the value calculated for
a (3,3) nanotube geometry, although the presence of (5,0)
nanotubes in the zeolite matrix cannot be discarded\cite{dia}.
It has been shown by using the local-density functional method
that the (3,3) nanotubes have the same band structure of typical
armchair nanotubes near the Fermi level, with a pair of
subbands crossing at two opposite momenta\cite{dia}. The 
case of the (5,0) nanotubes is different in that they have one more 
subband crossing the Fermi level, with angular momentum $L = 0$
in the zigzag geometry. We will show that the (5,0)
nanotubes have a extended low-energy phase
with strong charge-density-wave (CDW) correlations,
while the dominant low-energy instability for the (3,3) geometry
is given by the phase separation into regions with opposite
electronic current\cite{dme}. We will see that this is consistent
with the existence of a low-energy transition 
inherent to the individual CN, allowing also to account 
for the phenomenology reported in Ref. \onlinecite{chi}.

Focusing first on the (3,3) nanotubes, the interaction processes 
between different low-energy branches crossing at Fermi points 
$k_F$ and $-k_F$ can be classified by attaching to them respective 
coupling constants $g_i^{(j)}$\cite{berk}. The lower index discerns 
whether the
interacting particles shift from one Fermi point to the other
$(i=1)$, remain at different Fermi points $(i=2)$, or they
interact near the same Fermi point $(i=4)$. The upper label
follows the same rule to classify the different combinations of
left- and right-movers, including the possibility of
having Umklapp processes $(j=3)$ in the undoped system.

We will adopt in what follows a bosonization approach to deal 
with the interactions in forward-scattering (FS) channels, while 
the effect of the rest of interactions will be considered through 
the analysis of their low-energy scaling. Thus, we first introduce
an electron density operator $\rho_{ri} (k)$ for each of the 
linear branches of the subbands
crossing the Fermi level, where the index 
$r = L, R$ denotes the left- or right-moving character and 
the index $i=1, 2$ labels the Fermi point. It becomes possible to 
decouple the FS channels by passing to the combinations
\begin{equation}
\rho_{r \pm }(k) = 
\nu  ( \rho_{r 1 }(k) \pm \rho_{r 2 }(k) )
/ \sqrt{2}
\end{equation} 
where $\nu = -1,+1$ for $r = L,R$ , respectively.
The hamiltonian for the FS interactions can be written in terms
of fields $\Phi_{\pm } (x)$, such that
$\partial_x \Phi_{\pm }(x) = \rho_{L \pm }(x) + \rho_{R \pm }(x)$,
and their conjugate momenta $\Pi_{\pm } (x)$:
\begin{equation}
H_{FS}   =    \frac{1}{2}  \int dx
    \sum_{ s = \pm }  \left(   v_{Js} ( \Pi_{s} (x)  )^2  +
               v_{Ns} ( \partial_x \Phi_{s}(x) )^2      \right)
\end{equation}
The four independent velocities are given by
\begin{eqnarray}
v_{J\pm}=  v_F + (1/\pi ) \left( g_4^{(4)} \pm g_2^{(4)}
                - (g_2^{(2)} \pm g_4^{(2)} ) \right)
                                             \label{v2}  \\
v_{N\pm}=v_F +  (1/\pi )  \left( g_4^{(4)} \pm g_2^{(4)}
                + (g_2^{(2)} \pm g_4^{(2)} ) \right)
                                              \label{v1} 
\end{eqnarray}
where $v_F$ is the Fermi velocity.
In terms of these quantities, the renormalized velocities of 
the liquid are $u_{\pm}  =  \sqrt{v_{N\pm} v_{J \pm}}$, while
the charge stiffnesses are 
$K_{\pm}  =  \sqrt{ v_{J \pm} / v_{N \pm} }$ \cite{voit}.

In general, the significance of the interactions is given by
their scaling behavior at low energies. The scaling
equations of the $g_i^{(j)}$ couplings have been obtained in
Ref. \onlinecite{berk} for an armchair geometry. 
Here we improve nonperturbatively the equations
writing the scaling dimensions in terms of the $K_{\pm }$
parameters, and we introduce the pertinent modifications 
for the (5,0) geometry. The new equations read:
\begin{eqnarray}
\frac{\partial g_1^{(1)}}{\partial l}  & = &
    - \frac{1}{\pi v_F}  (  g_1^{(1)} g_1^{(1)}
        +  g_1^{(2)} g_2^{(1)} )
        - \frac{1}{\pi v_F} \beta u_F u_B   \label{first}   \\
\frac{ \partial g_1^{(2)}}{\partial l} & = &
     (1 - \frac{1}{K_{-}}) g_1^{(2)}
     +   \frac{1}{\pi v_F}  (  g_4^{(3)} g_1^{(3)}  \nonumber  \\
     & &   - g_2^{(1)} g_1^{(1)}  
            - (\beta /2)  ( u_F^2  + u_B^2 )  )         \\
\frac{ \partial g_2^{(1)}}{\partial l}  & = &
     (1 - \frac{1}{K_{-}})  g_2^{(1)}
   + \frac{1}{\pi v_F}  (  g_4^{(1)} g_1^{(2)} 
                     - 2  g_4^{(1)} g_2^{(1)}      \nonumber     \\
   &  &    + g_4^{(3)} g_1^{(3)} - g_4^{(3)} g_2^{(3)}  
       - g_1^{(2)} g_1^{(1)} - \beta u_F u_B )   \label{part}  \\
\frac{ \partial g_1^{(3)}}{\partial l}  & = &
        (1 - K_{+})  g_1^{(3)}
   + \frac{1}{\pi v_F}   (  - 2 g_1^{(3)} g_1^{(1)}   \nonumber \\
  &  &   +  g_2^{(3)} g_1^{(1)}  + g_4^{(3)} g_1^{(2)}  )     \\
\frac{ \partial g_2^{(3)}}{\partial l}  & = &
      (1 - K_{+}) g_2^{(3)}
  + \frac{1}{\pi v_F}  (  g_4^{(1)} g_1^{(3)}   \nonumber     \\
   &  &  - 2 g_4^{(1)} g_2^{(3)} + g_4^{(3)} g_1^{(2)}
                   - g_4^{(3)} g_2^{(1)}  )    \\
\frac{ \partial g_4^{(3)}}{\partial l}  & = &
      (2 - K_{+} - \frac{1}{K_{-}})  g_4^{(3)}
   +  \frac{1}{\pi v_F}   ( - g_4^{(3)} g_4^{(1)}   \nonumber   \\
    &  &   - 2  g_2^{(3)} g_2^{(1)} + g_1^{(3)} g_2^{(1)}
            + g_2^{(3)} g_1^{(2)} + g_1^{(3)} g_1^{(2)}   )
\label{last}
\end{eqnarray}
where $l$ stands for minus the logarithm of the energy
(temperature) scale measured in units of the high-energy scale
$E_c $ of the 1D model
(of the order of $\sim 0.1 \; {\rm eV}$).
The $g_i^{(3)}$ couplings only arise in the (3,3) nanotubes 
without doping (which is the likely experimental condition when
embedded in the insulating zeolite matrix) and the $u_F$ and 
$u_B$ couplings (to be defined later)
apply only to the (5,0) nanotubes.
The rest of equations not written here retain the expressions
given in Ref. \onlinecite{berk}.

The scaling equations have to be solved with initial conditions
accounting for the competition between the Coulomb interaction
and the effective interaction arising from 
phonon-exchange\cite{demler}. The latter turns out to be 
attractive in the backscattering (BS) channels 
and repulsive in the Umklapp channels\cite{th2,cara}. 
Its effective coupling $g$ has a strength $\approx 0.3 - 0.9$  
times $v_F / n$ for a $(n,n)$ nanotube\cite{cara}. This is 
comparable to the strength of the Coulomb interaction in BS and 
Umklapp processes, which we have taken as $\approx 0.2 \; e^2/ n$
following Refs. \onlinecite{eg} and \onlinecite{kane}. 
Given that $e^2 \approx 2.7 v_F$, the Coulomb 
repulsion only becomes dominant in forward-scattering processes
mediated by the potential $V (q) \approx (2e^2/ \kappa ) 
\log (1 + q_0/q)$, where $q$ is the momentum and $\kappa $ 
a suitable dielectric constant\cite{sarma}. Under these initial 
conditions, the scaling equations develop an unstable flow
below certain energy scale, at which the $g_2^{(1)}$ and
$g_4^{(1)}$ couplings enter a regime of large attraction while 
the $g_4^{(2)}$ and Umklapp couplings are driven towards large
repulsion in the (3,3) nanotubes.

\begin{figure}
\begin{center}
\mbox{\epsfxsize 4.5cm \epsfbox{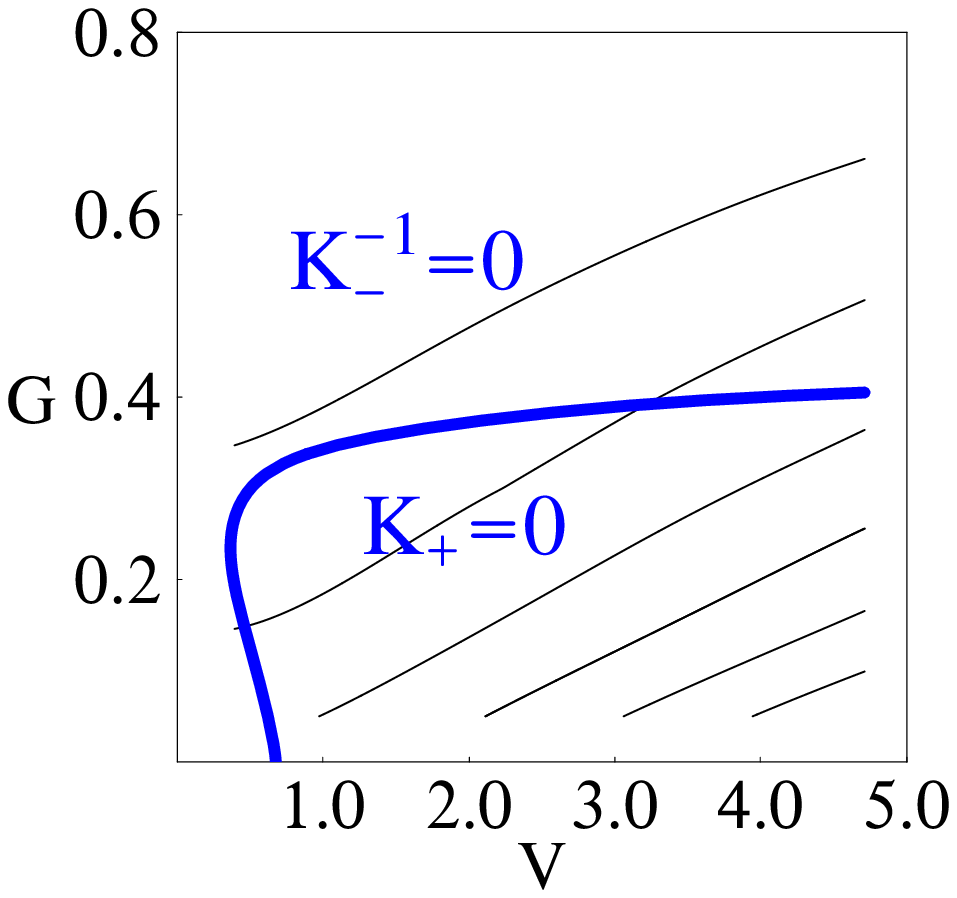}
  \epsfxsize 4.2cm \epsfbox{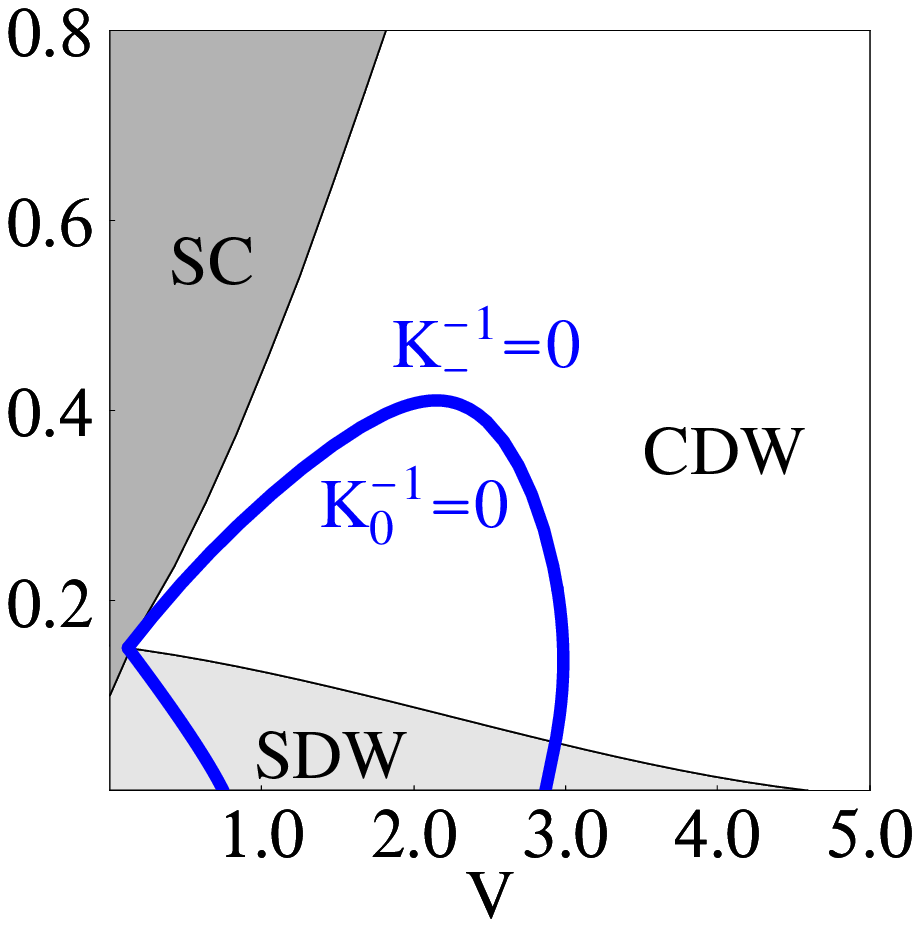}}
\end{center}
\caption{
Phase diagram of (3,3) (left) and (5,0) (right) CN,
in terms of the Coulomb potential $V$
(in units of $v_F$) and the effective coupling of
the phonon-exchange interaction $G = 4|g|/\pi v_F $.
The thick lines are boundaries between the different
phases characterized by the breaking of the parameters
$K_{\pm}$ and $K_0$ at low temperature scales. 
In the (3,3) nanotubes, CDW
correlations at zero momentum are dominant at the 
singular point, but with low strength represented 
by the contours of constant CDW response function 
$R = 2, 4, 8, 16, 32, 64$ (from top to bottom).
For the (5,0) nanotubes, three different regimes arise with
strong correlations at the singular point of $K_{-}$ or
$K_{0}$, given by the dominance of the $2k_F$ CDW 
(white area), the $2k_F$ spin-density-wave (light area),
or the $s$-wave SC response function (dark area).}
\label{one}
\end{figure}

We have discerned the character of the electronic instability
by looking for the susceptibility with the fastest growth as 
$l \rightarrow \infty$, as well as for the possible breakdown
of the Luttinger liquid parameters $K_{\pm}$. 
We have checked that this latter instance 
takes place before any CDW or pairing response function 
starts to diverge in the (3,3) nanotubes. Their phase diagram is
represented in Fig. \ref{one}. 
The strength of the Coulomb potential $V$ for the 
samples reported in Ref. \onlinecite{chi} has to be determined 
taking into account the screening effects from the 
large 3D array of CN inserted in the zeolite channels.
These effects can be studied by generalizing the many-body 
approach devised in Ref. \onlinecite{quinn}.
Thus, by performing a RPA analysis incorporating the
electrostatic coupling between all the CN in the array, we
have found that the Coulomb potential ranges from $V \approx
0.5 \; e^2$ at $q = 10 \; {\rm nm}^{-1}$ to a saturation value
$V \approx 0.9 \; e^2$ for 
$q = 0.1 \; {\rm nm}^{-1}$. Taking $e^2 \approx 2.7 v_F$ and
$|g| \approx 0.2 - 0.3 \; v_F$ for the
(3,3) nanotubes, we observe that the relevant regime corresponds
to the phase $K_{+} = 0$, with very weak SC or CDW correlations.

The vanishing of the $K_{+}$ parameter corresponds to an
instability in the balance between left- and right-moving 
electrons in the system. This follows from the fact that 
the vanishing of $v_{J+}$ in (\ref{v2}) is what 
triggers the instability. A phase with $K_{+} = 0$ has been 
already described in the study of CN with purely 
repulsive interactions\cite{yo}. This instance has been also
analyzed in Ref. \onlinecite{tsv}. In the case of CN of 
short radius, however, we have a nonnegligible
strength of the phonon-exchange interaction. This leads to a 
different physical situation as the velocity $v_{N+}$ 
remains finite, and the compressibility $K_{+}/u_{+}$ 
is nonvanishing at the point of the transition. 
We will see that the vanishing of $v_{J+}$ marks actually the 
divergence of the $\Pi_{+}$ correlations, opening a regime 
of phase separation into regions with opposite electronic 
current.

The onset of the phase characterized by the vanishing of 
$K_{+}$ explains the apparent gap that has been measured from 
the $I$-$V$ curves in Ref. \onlinecite{chi}. The 
development of the gap shown in Fig. 2 of
that Reference takes place through the change in
the functional form of the $I$-$V$ characteristics as the
temperature is lowered, ressembling the behavior
of a power-law dependence with increasing exponent. The
same kind of behavior has to be found in the
differential conductance $dI/dV$, which gives a measure of 
the density of states near the Fermi level.

\begin{figure}
\begin{center}
\mbox{\epsfxsize 6.5cm \epsfbox{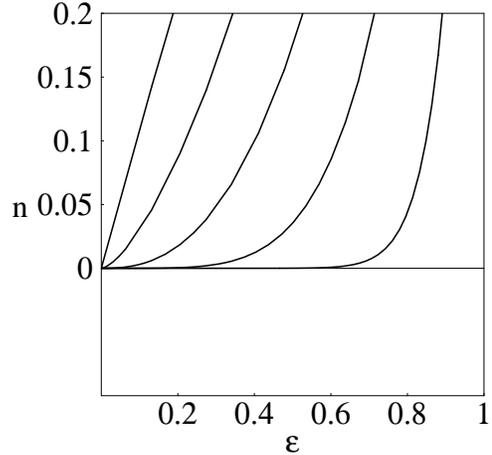}}
\end{center}
\caption{Plot of the density of states $n(\varepsilon )$
for $V = 0.75 \pi v_F$ and $4|g|/\pi v_F = 0.2$ . The curves
correspond to values of minus the logarithm
of the temperature scale $l = 3.21, 3.225, 3.232, 3.235$
and $3.236$ .}
\label{two}
\end{figure}

In our description, the depletion of the density of states 
$n(\varepsilon )$ near the Fermi level can be evaluated with 
the usual bosonization methods, with the result that
\begin{equation}
n(\varepsilon ) \sim \varepsilon^{(K_{+} + 1/K_{+} +
        K_{-} + 1/K_{-} - 4)/8}
\label{dos}
\end{equation}
The density of states follows
a power-law behavior with an increasingly large exponent as
$K_{+}$ vanishes in the low-temperature limit, as represented 
in Fig. \ref{two}. The plots in the figure show the development
of the pseudogap at low temperatures. We observe that
the shapes of the curves obtained from (\ref{dos}) 
are in very good agreement with the form of 
the $I$-$V$ characteristics reported in Ref. \onlinecite{chi}.

Shifting now to the case of the (5,0) nanotubes, their behavior
can be studied by making the parallel of the above analysis.
One has to incorporate an additional type of density 
operators for the subband with angular momentum $L = 0$ (and
Fermi velocity $v_F'$), which has its own renormalized
velocity $u_0$ and charge stiffness $K_0$. Moreover, there are
new interaction processes in which one or two of the incoming 
modes belong to the subband with $L = 0$. These 
interactions may be classified with an additional set of couplings 
$f_i^{(j)}$, where the indices $i$ and $j$ keep the same meaning 
as for the $g_i^{(j)}$ couplings. Finally, there may be processes 
in which two particles with opposite $L$ around the 
Fermi points 1 and 2 end up in the subband with $L = 0$, 
requiring the introduction of new couplings $u_F$ and 
$u_B$, for the respective cases with and without change of 
chirality of the particles.

The processes given by the couplings $u_F$ and $u_B$ have 
been overlooked in other analyses of the (5,0) 
nanotubes\cite{jap}, while they are 
determinant in the enhancement of the SC correlations. 
Their scaling equations are given by 
\begin{eqnarray}
\frac{ \partial u_F}{\partial l}  & = &    \Delta u_F   
       -  \frac{1}{2 \pi v_F}  ( g_1^{(2)} u_F
     + (g_2^{(1)}  +  g_1^{(1)} +  \beta f_4^{(1)}) u_B   )  \\
\frac{\partial u_B}{\partial l}  & = &    \Delta  u_B 
       -  \frac{1}{2 \pi v_F}  ( g_1^{(2)} u_B 
      + (g_2^{(1)}  +  g_1^{(1)}  + \beta f_4^{(1)}) u_F   )   
                                                \nonumber    \\
  &  &  +  \frac{\alpha }{\pi v_F} f_1^{(1)} (u_F - 2u_B)
\end{eqnarray}
where $\Delta = 1 - 1/4 K_{+} - 1/4 K_{-} - 1/2 K_0 + \alpha  
f_2^{(2)} /\pi v_F$, $\beta = v_F / v_F'$ and
$\alpha = 2/(1 + v_F'/v_F)$. The dominant contributions in these
channels come from phonon-exchange processes, so that the 
interactions $u_F$ and $u_B$ enter a regime of large attraction 
at low-energies. The flow of the scaling equations gives rise now 
to a fast growth of the $2k_F$ CDW, the $2k_F$ spin-density-wave,
or the $s$-wave SC response function, leading to the phases
shown in Fig. \ref{one}. Recalling the value of the couplings 
for the experimental samples, we see that the relevant region 
falls in this case into the phase with strong CDW correlations.

An important point in connection with the phenomenology 
reported in Ref. \onlinecite{chi} is that the (3,3) nanotubes 
fall at low energies into a phase which is consistent with the 
large diamagnetic signal observed in the experiments. This 
can be seen by recalling that the vector part ${\bf A}$ of 
the electromagnetic potential couples to the electronic current. 
Actually, around each Fermi point, the 
modes of the electron field can be arranged into a bispinor 
$\Psi $, whose hamiltonian is obtained from that of the 
low-energy excitations of graphene\cite{dirac}:
\begin{equation}
  H_G  = -i  v_F \int d^{2} r \; \Psi^{\dagger} (\mbox{\bf r})
\mbox{\boldmath $\sigma \cdot$} 
( \mbox{\boldmath  $\nabla $} - i (e/c) {\bf A} ) 
                        \Psi(\mbox{\bf r})
\label{ham}
\end{equation}
The 1D projection to the low-energy excitations of the
nanotube can be done consistently in the case of a transverse
magnetic field, by chosing the vector potential as usual in 
the longitudinal direction of the nanotube\cite{dress}. After 
diagonalizing the quadratic form in (\ref{ham}), the 1D 
hamiltonian becomes
\begin{eqnarray}
  H_{1D}  & = &  v_F \int d k d \varphi       
\; (k - (e/c)  A_{\parallel} (\varphi ) )      \nonumber      \\
   &    &   \;\;\;\;\;\;\;\;\;\;\;
( \Psi^{\dagger}_R(k) \Psi_R(k) - \Psi^{\dagger}_L(k) \Psi_L(k) )
\label{ham1D}
\end{eqnarray}
where $\varphi $ is the angle that supports the modulation of 
the longitudinal component $A_{\parallel}$ around the nanotube.
It becomes clear that the vector potential couples to the 
charge asymmetry $\Pi_{+} = \rho_{R 1} - \rho_{L 1} + \rho_{R 2} 
- \rho_{L 2}$.

When there is an enhanced susceptibility for the charge
mismatch between left and right branches, a mechanism of
screening of the magnetic field takes place, similar to the 
usual screening of the electric field from the total electron
charge. The response function for the current 
$\Pi_{+}$, computed at the bosonization level, is
\begin{equation}
\langle \Pi_{+}(\omega , k ) \Pi_{+}(-\omega , -k ) \rangle
= \frac{1}{K_{+}} \frac{u_{+} k^2}{\omega^2 - u_{+}^2 k^2}
\label{sus}
\end{equation}
After integration of the density fields, we obtain a 
contribution to the free-energy density which, in the static
limit, has the form   $ - A_{\parallel}(\varphi ) 
A_{\parallel}(\varphi ) \; v_F^2 (e/c)^2 /K_{+}u_{+} $.  
The susceptibility (\ref{sus}) leads therefore to a large 
diamagnetic response in the regime where $1/K_{+}u_{+}$
diverges.

The instability in the current is triggered by 
the vanishing of the renormalized velocity $v_{J+}$ 
at low energies. The diamagnetic response is
therefore enhanced by the vanishing of $K_{+}$ and 
$u_{+}$ at low temperatures. The behavior of
the susceptibility $v_F^2 (e/c)^2 /K_{+}u_{+}$ as a 
function of the temperature is actually in qualitative 
agreement with the divergence of the magnetic susceptibility 
of CN reported in Ref. \onlinecite{chi}.

We end up with a picture consistent with the experimental
observations reported in Ref. \onlinecite{chi}, assuming that,
as supported by the measurements of the nanotube diameter, 
most part of the CN contained
in the zeolite matrix have preferently a (3,3) geometry.
We have seen that the transition reported in Ref.
\onlinecite{chi} may be interpreted as the breakdown of the
Luttinger liquid regime, originated from
the onset of phase separation into regions with opposite
electronic current.
As well as for graphene\cite{dress}, the existence
of degenerate points in the spectrum is at the origin of the 
diamagnetic signal in the CN. The peculiar feature
in the nanotubes is that their 1D character leads to the
splitting of the electron system into regions with alternating
direction of the current. This implies necessarily the
appearance of domain walls in the configuration of the current. 
The physical picture provides then a 1D analogue of a
well-known phenomenon in higher dimensions, by which the
disordering effect of solitons spoils the propagation of
the gauge fields. This is the role of the divergent
fluctuations and consequent domain walls in the current, 
which lead to a phase transition in 1D as that only requires 
the breakdown of a discrete symmetry, namely that of the 
current direction in the CN.



\begin{references}


\bibitem{bal}
L. Balents and M. P. A. Fisher, Phys. Rev. B {\bf 55}, R11973
(1997).

\bibitem{eg}
R. Egger and A. O. Gogolin, Phys. Rev. Lett. {\bf 79}, 5082
(1997); Eur. Phys. J. B {\bf 3}, 281 (1998).

\bibitem{kane}
C. Kane, L. Balents and M. P. A. Fisher, Phys. Rev. Lett. {\bf
79}, 5086 (1997).

\bibitem{yo}
H. Yoshioka and A. A. Odintsov, Phys. Rev. Lett. {\bf 82}, 374
(1999); Phys. Rev. B {\bf 59}, R10457 (1999).


\bibitem{exp}
M. Bockrath {\em et al.}, Nature {\bf 397}, 598 (1999).  

\bibitem{yao}
Z. Yao {\em et al.}, Nature {\bf 402}, 273 (1999).

\bibitem{kas}
A. Yu. Kasumov {\em et al.}, Science {\bf 284}, 1508 (1999).

\bibitem{marc}
A. F. Morpurgo {\em et al.}, Science {\bf 286}, 263 (1999).

\bibitem{th1}
J. Gonz\'alez, Phys. Rev. Lett. {\bf 87}, 136401 (2001).

\bibitem{sup} 
M. Kociak {\em et al.}, Phys. Rev. Lett. {\bf 86}, 2416 (2001).
A. Kasumov {\em et al.}, Phys. Rev. B {\bf 68}, 214521 (2003).

\bibitem{chi}
Z. K. Tang {\em et al.}, Science {\bf 292}, 2462 (2001).

\bibitem{th2}
J. Gonz\'alez, Phys. Rev. Lett. {\bf 88}, 076403 (2002); 
Phys. Rev. B {\bf 67}, 014528 (2003).

\bibitem{th3}
J. V. Alvarez and J. Gonz\'alez, Phys. Rev. Lett. {\bf 91}, 
076401 (2003).

\bibitem{dia}
H. J. Liu and C. T. Chan, Phys. Rev. B {\bf 66}, 115416 (2002).

\bibitem{dme}
Phase separation has been also advocated in CN with dominant
attractive interaction by
A. De Martino and R. Egger, Phys. Rev. B {\bf 67}, 235418 (2003).


\bibitem{berk}
Yu. A. Krotov, D.-H. Lee and S. G. Louie, Phys. Rev. Lett.
{\bf 78}, 4245 (1997).

\bibitem{voit}
J. Voit, Rep. Prog. Phys. {\bf 58}, 977 (1995).

\bibitem{demler}
The out-of-plane optical modes play a dominant role in 
CN of short radius, followed by in-plane optical modes 
and breathing modes, according to R. Barnett, E. Demler and
E. Kaxiras, report cond-mat/0305006.

\bibitem{cara}
A. S\'ed\'eki, L. G. Caron and C. Bourbonnais, Phys. Rev. B
{\bf 65}, 140515 (2002).

\bibitem{sarma}
D. W. Wang, A. J. Millis and S. Das Sarma, Phys. Rev. B {\bf 64},
193307 (2001).

\bibitem{quinn}
P. Hawrylak, G. Eliasson and J. J. Quinn, Phys. Rev. B {\bf 37},
10187 (1988). 

\bibitem{tsv}
A. A. Nersesyan and A. M. Tsvelik, Phys. Rev. B {\bf 68},
235419 (2003).

\bibitem{jap}
K. Kamide {\em et al.}, Phys. Rev. B {\bf 68}, 024506 (2003).

\bibitem{dirac}
J. Gonz\'alez, F. Guinea and M. A. H. Vozmediano, Nucl. Phys. B 
{\bf 424}, 595 (1994).

\bibitem{dress}
R. Saito, G. Dresselhaus and M. S. Dresselhaus,
{\em Physical Properties of Carbon Nanotubes}, Chap. 6,
Imperial College Press, London (1998).






\end{references}
\end{document}